# Electron Spin Dephasing and Optical Pumping of Nuclear Spins in GaN

G. Wang[1,2], C. R. Zhu[1], B. L. Liu[1*], H. Ye[1], A. Balocchi[2], T. Amand[2], B. Urbaszek[2], H. Yang[3] and X. Marie[2**]

[1]*Beijing National Laboratory for Condensed Matter Physics, Institute of Physics, Chinese Academy of Sciences, Beijing 100190, People's Republic of China*

[2]*Université de Toulouse, INSA-CNRS-UPS, LPCNO, 135 Avenue de Rangueil, 31077 Toulouse, France*

[3]*Suzhou Institute of Nanotech and Nanobionics, Chinese Academy of Sciences, Suzhou 215123, People's Republic of China*

*We have measured the donor-bound electron spin dynamics in cubic GaN by time-resolved Kerr rotation experiments. The ensemble electron spin dephasing time in this quantum dot like system characterized by a Bohr radius of 2.5 nm is of the order of 1.5 ns as a result of the interaction with the fluctuating nuclear spins. It increases drastically when an external magnetic field as small as 10 mT is applied. We extract a dispersion of the nuclear hyperfine field $\delta B_n \sim 4\,mT$, in agreement with calculations. We also demonstrate for the first time in GaN based systems the optical pumping of nuclear spin yielding the build-up of a significant nuclear polarization.*

A localized electronic spin in a semiconductor is a promising candidate for implementing quantum information processing in solids [1-6]. The confinement of single electrons in (In)GaAs semiconductor quantum dots or on neutral donor atoms $D_0$ in GaAs yields long coherence times at low temperature which allows the realization of quantum optical control operations. In both systems the hyperfine interaction between the electron and nuclear spin plays a crucial role. On one hand the electron spin dephasing time is limited by the fluctuation of the surrounding nuclear spin which can be an obstacle for achieving coherent electron spin control [7]. On the other hand the Dynamic Nuclear Polarization (DNP) which corresponds to the transfer of electron spin to nuclear spins yields the build-up of an effective nuclear polarization which can also be used to store quantum information with very small coupling to the environment.

Though wide band gap GaN-based semiconductors are nowadays key materials for the electronics and optoelectronics industry, very few measurements of the carrier spin properties have been performed compared to GaAs-based structures [8]. Gallium nitride is potentially interesting for spintronics thanks to its smaller spin-orbit interaction compared to GaAs; it can crystallize in either the wurtzite (Wz) or zinc-blende cubic (ZB) structure. In Wz (In)GaN structures, electron and hole spin relaxation times of a few hundreds of ps [8-11] have been measured and even shorter times (~1 ps) were found for exciton spin depolarization times [12,13]. In Wz nitride nanostructures (quantum wells or quantum dots), exciton spin relaxation times of the order of 200 ps were reported at $T$ =300 K [14]. In these Wz nitride structures, the electronic and spin properties can be highly affected by the strong built-in electric field due to the spontaneous and piezoelectric polarizations [15]. In contrast these polarizations are negligible in ZB GaN structures or small size Wz quantum dots and much longer carrier spin relaxation times can be expected [16]. Electron spin relaxation times of 600 ps in highly doped bulk ZB cubic GaN (n~ $10^{19}$ cm$^{-3}$) [17] and exciton spin relaxation times longer than 10 ns in GaN quantum dots were indeed reported recently at room temperature [18].

Advantageously and contrary to the quantum dots characterized by strong fluctuations (in size, composition or strain), all the donor electrons in a semiconductor sit in the same environment and have the same wavefunction. The wave function of a neutral donor-bound electron ($D_0$) in GaN is well described by a hydrogenic wavefunction with a Bohr radius of $a_B$~ 2.5 nm (i.e. 4 times smaller than in GaAs) [19]. The donor electrons in GaN are also characterized by a much larger binding energy (~25 meV) [20, 21] compared to the one in GaAs (~5 meV). The electron spins will thus be more isolated from possible charge fluctuations occurring in the conduction band and make them good candidates for semiconductor spin qubits. The larger $D_0$ binding energies compared to GaAs should also allow higher temperature operation. However the donor electron spin dynamics in GaN has never been studied so far. Moreover each atom in GaN carries a nuclear spin (I=3/2 for Ga and I=1 for N). Surprisingly the effect of the hyperfine interaction on the electron spin properties in this material has not been evidenced so far either [22].

In this Rapid Communication we demonstrate that the spin properties of donor electrons in cubic GaN are governed by the interaction with the surrounding nuclear spins. We show that the electron spin relaxation time of donor-bound electrons, which is about 1.5 ns, increases strongly when an external magnetic field as small as 10 mT is applied. This demonstrates that the electron spin relaxation time is controlled by the spin dephasing of electrons due to the fluctuating nuclear spins, which is screened by the application of a small external field. For the sake of simplicity, the spin dephasing mechanism on the donor electrons ensemble is termed here ''spin relaxation''. We measured a fluctuation dispersion of the nuclear hyperfine field of the donor-bound electrons of about $\delta B_n \sim 4$ mT, in agreement with calculations. In addition we evidence for the first time in GaN materials the optically induced Dynamical Nuclear Polarization (DNP) which could be useful in a wide range of nuclear magnetic resonance (NMR) and magnetic resonance imaging experiments [23].

The cubic GaN epilayer investigated here has a thickness of 0.6 μm. It has been grown on GaAs (001) substrates by metal-organic vapor phase epitaxy (MOVPE) ; details on growth conditions can be found in [21]. Based on the measured resistivity, the n-type background doping is estimated to be in the range $10^{13}$-$10^{14}$ cm$^{-3}$ [21]. Photoluminescence (PL) spectroscopy has been performed in order to characterize the crystal quality of the sample. At $T$=10 K, two main lines are recorded at 3.271 and 3.155 eV respectively (figure 1a). These peaks coincide exactly with the ones measured by previous groups and correspond respectively to bound exciton and Donor-Acceptor-Pair (DAP) transitions [20, 21]. Figure 1b displays the temperature dependence of the exciton line energy. The dotted line represents the calculated free exciton energy as a function of temperature, assuming an exciton binding energy $E_b$=24 meV and the band gap variation law $E_g(T) = 3,302 - 6.7 \times 10^{-4} T^2 / (T+600)$ [20,21]. Above 100 K, we note an excellent agreement between the measured PL peak and the calculated free exciton energy. Below 100 K the measured exciton peak is smaller than the calculated free exciton energy due to the contribution of bound exciton transition as previously observed [20] (the donor bound exciton $D_0X$ and free exciton X lines are not resolved). Whatever the temperature is, no trace of near-band edge or defect related emission lines from inclusions of hexagonal crystal structure are found. This result, in agreement with structural characterization [21], confirms the high quality of this cubic GaN sample.

For the pump-probe measurements the sample is excited by a mode-locked frequency doubled Ti:sapphire (pulse duration~120fs, repetition rate : 76 MHz). The laser beams are focused to a ~100μm spot. The pump (circularly polarized σ+ or σ-) and the probe (linearly polarized) beams have an average power of 10 and 1mW, respectively. The circularly polarized pump pulse incident normal to the sample creates spin-polarized electrons with the spin vector directed along the growth direction of the sample. The electron spin dynamics is studied by Time Resolved Kerr Rotation (TRKR) where the pump-induced change of the linearly polarized polarization beam is detected as a function of the delay time between the pump and the probe. After reflection on the sample, the probe beam is decomposed into its two linear components, and the difference in their intensities is measured with a balanced optical bridge [24]. The laser wavelength is set at the maximum Kerr rotation signal which follows very well the peak energy of exciton photoluminescence (see figure 1.b). The experiments are carried out in an Oxford magneto-optical cryostat supplied with a 7-T split-coil super-conducting magnet. All the experiments presented below were performed at T=2 K.

In the first series of experiments, the pump beam helicity (σ+, σ-) is modulated at 50 kHz with a photo-elastic modulator (PEM) coupled to the lock-in detection system. Because of the very slow nuclear spin dynamics these experimental conditions prevent the built-up of a nuclear spin polarization through optically induced dynamical nuclear polarization [5]. Figure 2 shows the transient Kerr rotation signal for different values of the magnetic field $B_z$ applied parallel to the excitation light propagation direction (Faraday configuration). For $B_z$=0, the decay time of the Kerr signal is of the order of 1500 ps (non mono-exponential); it corresponds to the time evolution of the average electron spin polarization <$S_z$> of the electrons which have been spin polarized by the circularly polarized pump beam (z is parallel to the excitation light direction and perpendicular to the sample plane) [25]. Note that the initial decay time of the Kerr signal for t<~500 ps is controlled by electron-hole recombination [8]. The application of an external magnetic field as small as $B_z$=10 mT yields a spectacular increase of the Kerr signal decay time. We emphasize that for such a weak external field the electron Zeeman splitting $g_e\mu_B B$ is more than one hundred time smaller than $k_B T$; here $g_e$ is the electron effective Lande factor, $\mu_B$ is the Bohr magneton and $k_B$ the Boltzman constant (using $g_e$ =1.93±0.02 as measured in the second part of this letter from spin quantum beats experiments). In the presence of the external magnetic field we observe in figure 2b a non-zero Kerr rotation signal at negative pump-probe delays, indicating that the donor electron spin polarization is not fully relaxed within the 13 ns repetition period of the laser pulses.

For the very low doping concentration in this cubic GaN sample, the average distance between donors is of the order of 200 nm. Thus at low temperatures, the electrons are localized at the donor atoms and following the results obtained in GaAs it is a good approximation to neglect the electron-electron interactions [26,27]. Because of the strong localization of the donor electron hydrogenic wavefunction, the classical spin relaxation mechanisms based on spin-orbit interaction well known for bulk non-centrosymmetric materials, such as the D'Yakonov-Perel one, are suppressed in a similar way as in quantum dots and the main electron spin dephasing time is due to the interaction with the fluctuating nuclear field [5].

The hyperfine interaction of a localized electron spin with the surrounding nuclei can be described in a mean field approach by a frozen effective nuclear field acting on the electron spin [28]. The nuclear field mean value $\mathbf{B}_n = \langle \mathbf{B}_n \rangle$ fluctuates from donor to donor due to different realizations of the nuclear spins configuration. The dispersion of this nuclear hyperfine field $B_n$ in the absence of dynamic nuclear polarization can be described by a Gaussian distribution $W(B_n) \propto \exp(-3B_n^2 / 2\delta B_n^2)$, where the fluctuations (root mean square deviation) are described by an effective field $\delta B_n = \sqrt{\langle B_n^2 \rangle - \langle B_n \rangle^2}$. This frozen fluctuation approach is justified since the correlation time of the nuclear field $\mathbf{B}_n$ ($10^{-4}$ s) is several orders of magnitude longer than the typical electron-spin dephasing time [28]. Repeated measurements of the expectation value of $\mathbf{B}_n$ at time intervals longer than the nuclear spin correlation time give an average $\langle \mathbf{B}_n \rangle_{av} = 0$ under zero applied external magnetic field so that $\delta B_n = \sqrt{\langle B_n^2 \rangle}$. However each donor electron spin precesses coherently around the effective nuclear magnetic field $\delta B_n$ (inset in figure 2a). The average electron spin polarization in the donor ensemble thus decays with time because of the random distribution of the local nuclear effective fields.

In an external magnetic field applied along the z direction, the effect of the hyperfine interaction on the electron spin polarization along z can be strongly reduced if the amplitude of the external field $B_z$ is larger than the dispersion of the sample in-plane fluctuations of the nuclear hyperfine field. Figure 2c shows the Kerr signal measured at negative delay ($t=-60$ ps) obtained for several values of the longitudinal applied magnetic field. The observed increase of the measured Kerr signal at $t=-60$ ps (i.e., 13 ns after the previous pump pulse) when $B_z$ increases reflects a significant increase of the donor electron spin polarization [29-31]. We measure a characteristic Half Width at Half Maximum (HWHM) $\delta B_n = 4.5 \pm 1$ mT. This drastic increase of the electron spin polarization in weak external longitudinal magnetic field is a fingerprint of the electron spin dephasing induced by the interaction with nuclear spins [5,32].

To confirm this interpretation, we have calculated the spin dephasing time $T_\Delta$ of GaN donor electron due to the Fermi-contact hyperfine interaction. This dephasing time writes [28]:

$$T_\Delta = \hbar \sqrt{\frac{3N_L}{2n \sum_{j=1}^{n} I^j(I^j+1) \cdot (A^j)^2}} \quad (1)$$

where $N_L$ is the number of nuclei interacting with the donor electron, $A^j$ the hyperfine constant, $I^j$ the spin of the jth nucleus and $n$ is the number of nuclei per unit cell. The sum goes over all the atoms in the primitive unit cell. We take for the hyperfine constants of the Gallium atom the average value for the two isotopes $^{69}$Ga and $^{71}$Ga : $A^{3/2}$=42 μeV [5]. Due to the larger gyromagnetic ratios (factor 3.5) and spin (factor 1.5) of Ga compared to the ones of N, the hyperfine interaction with the Ga nuclei strongly dominates and the effects of the nitrogen nuclei can be neglected to a first approximation [33]. Using a Bohr radius of $a_B$=2.5 nm for the donor electron, we estimate that the number of nuclei in interaction with the donor electron is $N_L \sim 4.10^4$ [34]. Equation (1) yields: $T_\Delta \sim$1300 ps, in rather good agreement with the experimental value.

From the measured decay time of the Kerr signal for $B_z$=0 in figure 2a (~1500 ps) we can estimate the dispersion of the nuclear hyperfine field. We find $\delta B_n = \hbar/(g_e \mu_B T_\Delta) \sim 3.5 \pm 1.0$ mT. This value is perfectly consistent with the measured half width at half maximum in figure 2c. However, the dispersion extracted from this magnetic field experiment should take into account possible effects of the periodic excitations on the Kerr rotation signal (see Testelin et al. [35]). For $B_z$ as small as 10 mT, about 3 times larger than $\delta B_n$, we observe clearly the quenching of the spin relaxation induced by nuclei. Spin Echo or mode-locking experiments should allow the determination of the long donor electron coherence time $T_2$ [36].

In the second part of this Letter, we demonstrate the optical pumping of nuclear spins which was never observed in GaN structures to the best of our knowledge. The dynamic nuclear polarization in semiconductor results from the scalar form of the Fermi-contact hyperfine interaction which conserves the total spin [5]. When an electron spin, which has been initialized optically, relaxes its initial orientation via this interaction, its spin angular momentum is transferred to the nuclear spins. Since the nuclei are much less coupled to the lattice, their polarization relaxes on a much longer time scale than the electron spin polarization and a large DNP degree can in principle be reached. This will result in the build-up of a significant nuclear field $\mathbf{B}_n = \Omega_0 \sum_j A^j |F_{1s}(r^j)|^2 \langle \mathbf{I}^j \rangle / g_e \mu_B$, ($F_{1s}(r^j)$ is the donor hydrogenic envelope function at nucleus $j$, $\Omega$ is the volume of the elementary cell) which in turn acts back on the electron spin, *ie.* the so called Overhauser effect. The optically induced nuclear polarization can thus be probed by monitoring the modifications of the electron spin polarization dynamics which will depend on both the external magnetic field **B** and the nuclear field $\mathbf{B}_n$. This nuclear hyperpolarization which can be optically created could be very useful to enhance NMR signals for various applications [33].

In order to measure the optically induced nuclear field, an external magnetic field is applied at 45° with respect to the sample plane as shown in the inset of figure 3 [34,37]. Following circularly polarized light pulsed excitation propagating along the z direction, the initial electron spin polarization $S(0)$ is aligned along z. The external magnetic field causes these spins to precess (Larmor precession). However a non-precessing component of electron spin, $\mathbf{S}_{av}$, parallel to the field **B** remains. Integrated over many laser pulses, a significant average nuclear spin <**I**> parallel to $S_{av}$ will build-up through mutual spin flip-flops with lattice nuclei driven by the hyperfine interaction. The average nuclear spin $\langle \mathbf{I} \rangle$ reacts back on the electron spin as an effective (Overhauser) magnetic field given simply by $\mathbf{B}_n \approx f B_n^{max} \mathbf{S}_{av}$. Here $f$ is the nuclear spin leakage factor and $B_n^{max}$ is the maximum nuclear field [34]. Since, as stated before, the nitrogen hyperfine constant is small with respect to the one of Ga isotopes, $B_n^{max}$ is proportional to the average hyperfine constant $A$ for the $^{69}$Ga and $^{71}$Ga isotopes [38]. Note that the direction of the nuclear field $\boldsymbol{B}_n$ is determined by the signs of both the nuclear and the electron g-factors. As the donor electrons $g$ factor is positive [19], $\boldsymbol{B}_n$ and $\boldsymbol{S}_{av}$ have the same direction.

The optically induced nuclear field is measured through the variation of the electron Larmor precession frequency $\omega$ under different experimental conditions. First the electron spin dynamics probed by TRKR is recorded in a tilted external magnetic field $B$=1 T using a modulation of the circularly polarized pump excitation at a frequency of 50 kHz. This modulation of the excitation light polarization prevents the built-up of the nuclear spin polarization because of the very slow dynamics of the nuclear spins [5]. We observe in figure 3a the expected oscillations of the Kerr signal resulting from the Larmor precession of the electron spin. We measure a Larmor precession frequency $\omega$ = 170.6 GHz. From the linear dependence of the precession frequency with the external field (inset in figure 3a), we measure a donor electron g factor $g_e=1.93\pm0.02$. As expected the electron spin dynamics is not changed if the direction of the magnetic field is reversed confirming the absence of any Overhauser effect in these experimental conditions. Note that the Kerr rotation signal is always positive *i.e.* it is not symmetrical with respect to the time-axis. This is a direct consequence of the 45° tilted magnetic field configuration. If we neglect the spin relaxation processes, the z component of the average electron spin, which is probed by the Kerr signal, writes simply:

$$<S(t)> = -\frac{1}{2}\left(1-\left(1-\cos(\omega t)\right)/2\right) \qquad (2)$$

The modulation amplitude is reduced by a factor 2 compared to the Voigt configuration, in agreement with the measurement in figure 3a.

Figure 3b displays the result of the same experiment performed with no modulation of the polarized excitation pump (i.e. fixed σ+ polarized light). Remarkably we see that the measured precession frequency is no more the same for $+B$ or $-B$ applied external field (for a fixed pump σ+ polarization). We measure ω= 86.5 GHz for $B$=+0.5 T and ω= 83.5 GHz for B=-0.5 T. Note the clear temporal shift between the two curves in figure 3b. Due to the build-up of the nuclear polarization the Larmor precession frequency writes : $\omega = g_e\mu_B|B+B_n|/\hbar$ for the external field $+B$ and $\omega = g_e\mu_B|-B+B_n|/\hbar$ for the external field $-B$. From the Larmor precession frequency changes we get an Overhauser field $B_n = 9 \pm 0.5$ mT. This Overhauser field does not change if the external field varies from 0.5 to 2 T (not shown).

The interpretation of the observed shift in figure 3b in terms of a nuclear polarization effect is further confirmed by three complementary experiments:
(i) As expected the optically induced nuclear field $B_n$ decreases if we reduce the average optical excitation power (not shown) [37].
(ii) The Overhauser field decreases down to $B_n$~3 mT at $T$=40 K and completely vanishes at higher temperature as a result of the ionization of the donor electron (see figure 1b).
(iii) We measure the same temporal shift (yielding the same $B_n$) as in figure 3b when the direction of the magnetic field is fixed but we record the electron spin quantum beats following either σ+ or σ- polarized pump light. This demonstrates that the thermal electron spin polarization (the equilibrium electron spin polarization due to applied magnetic field) yields a negligible nuclear spin polarization; thus the nuclear field evidenced in figure 3b has been optically induced.

For uniform nuclear polarization, the field $B_n$ is independent of the electron localization volume [5,34] as the maximum Overhauser shift writes simply $g_e\mu_B B_N^{max} = I^{Ga}A^{Ga} + I^N A^N \sim 60$ μeV. This corresponds to $B_n^{max} \sim 530$ mT for fully polarized nuclei in GaN. Thus the measured Overhauser shift in figure 3b corresponds to a nuclear polarization $<I>/I$ of about 2 %. Though the nitrogen contribution to the total Overhauser field is almost negligible, this nuclear polarization value is close to the one obtained in similar experimental conditions in slightly n-doped GaAs [39]. Future experiments will allow the determination of the built-up time of the nuclear polarization and the characteristic nuclear spin diffusion length in GaN [37]. Finally we emphasize that the bound electron spin on donor in GaN can form a spin-qubit system with a strong radiative coupling to the bound exciton state with relatively small inhomogeneous broadening of the optical transitions. So these results also pave the way to coherent control of light using techniques based on electromagnetically induced transparency in GaN which can find applications for linear optics quantum computation and in the creation of large optical nonlinearities for photonic gates in nonlinear optics quantum computation [40,41].

In summary, we have evidenced that the donor bound electron spin dephasing time in GaN is controlled by the hyperfine interaction with the surrounding fluctuating nuclear field. A drastic enhancement of the spin relaxation time is measured when an external magnetic field as small as 10 mT is applied. In addition we have demonstrated the optical pumping of nuclear spin in GaN. This result opens the route to all optical nuclear magnetic resonance in semiconductor nitride systems.

*Acknowledgments :*

This work was supported by the National Science Foundation of China Grant No. 11174338 and the joint France-China ANR-NSFC project "SpinMan". X.M. acknowledges the support by the Chinese Academy of Sciences Visiting Professorship program for Senior International Scientists. Grant No. 2011T1J37.

**Figure Caption**

**Figure 1**
(a) Photoluminescence (PL) spectrum of the cubic GaN at $T$=10 K evidencing both the exciton transition and the Donor-Acceptor Pair (DAP) transition; the laser excitation energy is $E_{laser}$=3.815 eV ; inset : PL spectrum at room temperature.
(b) Temperature dependence of the exciton PL peak energy (circle), Kerr rotation maximum signal energy (square). The solid line corresponds to the gap energy ($E_g$) variation and the dotted curve is the calculated free exciton energy transition assuming an exciton binding energy of 24 meV (see text).

**Figure 2**
T=2 K (a) Normalized Kerr Rotation Dynamics for external magnetic fields $B_z$=0 and 10 mT ; inset : schematics of the donor electron spins ensemble in interaction with the surrounding nuclei characterized by a fluctuation hyperfine field $\delta B_n$ (red arrows) and the external field $B_z$.
(b) Kerr rotation dynamics (raw data) highlighting the non-zero signal at negative delays due to the enlarged electron spin dephasing time in the external magnetic field $B_z$.
(c) Magnetic field dependence of the TRKR signal at negative delay ($t$=-60 ps) yielding a Half Width at Half Maximum of 4.5± 1 mT. The line is a guide to the eyes.

**Figure 3**
T=2 K ; Tilted magnetic field configuration (45°)
(a) Modulated (σ+/σ−) excitation. Kerr rotation dynamics for $B$=+1 T and $B$=-1 T. Inset: Larmor precession frequency ω as a function of the external magnetic field B.
(b) Fixed σ+ excitation. Kerr rotation dynamics for $B$=+0.5 and $B$=-0.5 highlighting the precession frequency changes due to the Overhauser effect (see text).
Inset : Schematics of the electron spin dynamics in a positive magnetic field +B for the σ+ case of the modulated (σ+/σ-) excitation (top) or a fixed σ+ excitation (bottom) . The latter evidences the optical pumping of nuclear spin polarisation and the associated Overhauser field $B_n$.

\* blliu@iphy.ac.cn
\*\* marie@insa-toulouse.fr

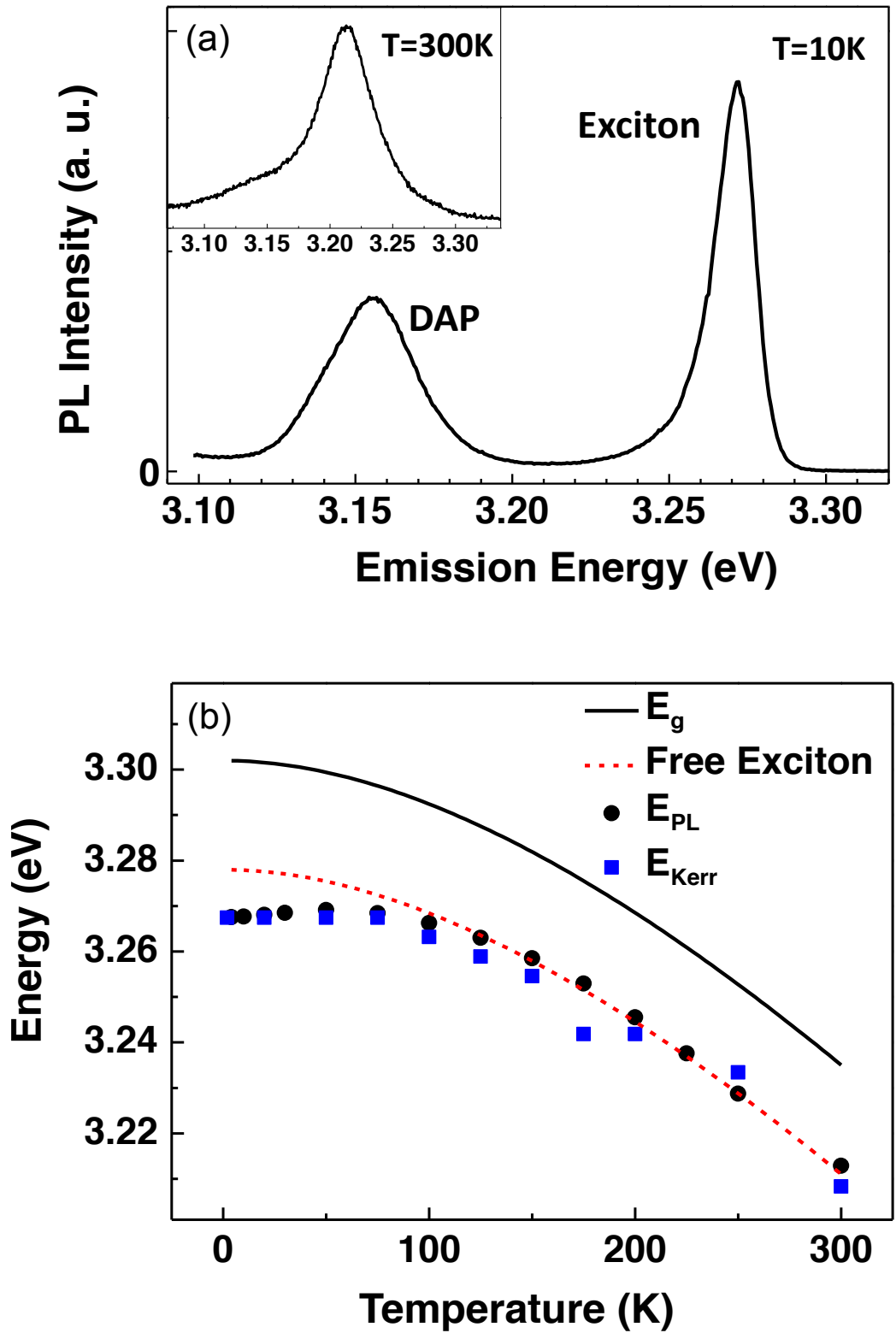

**Figure 1**

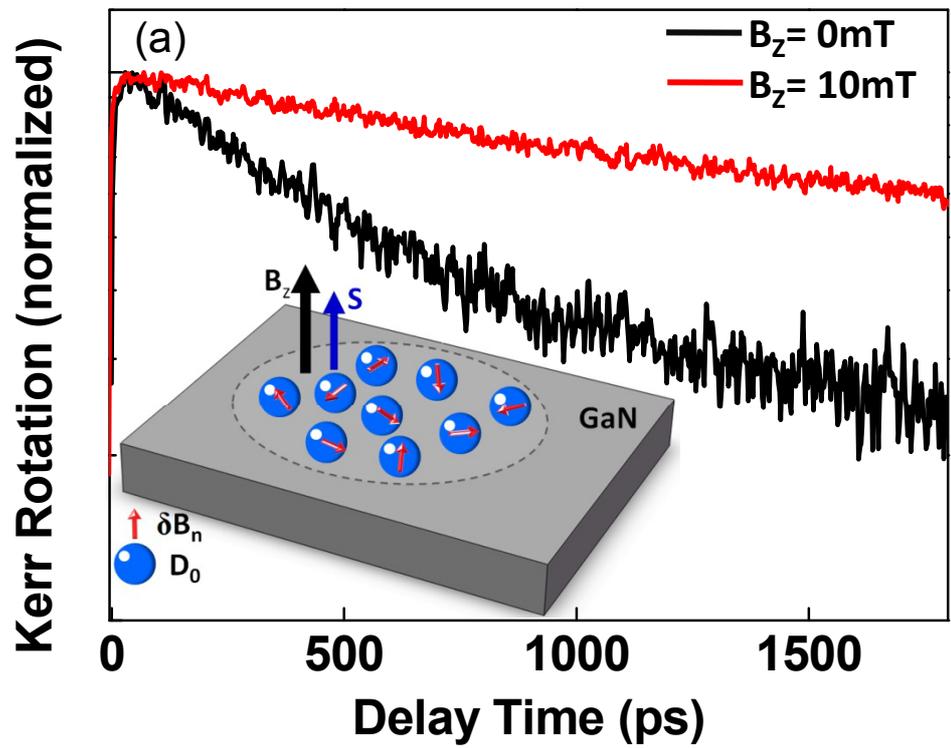

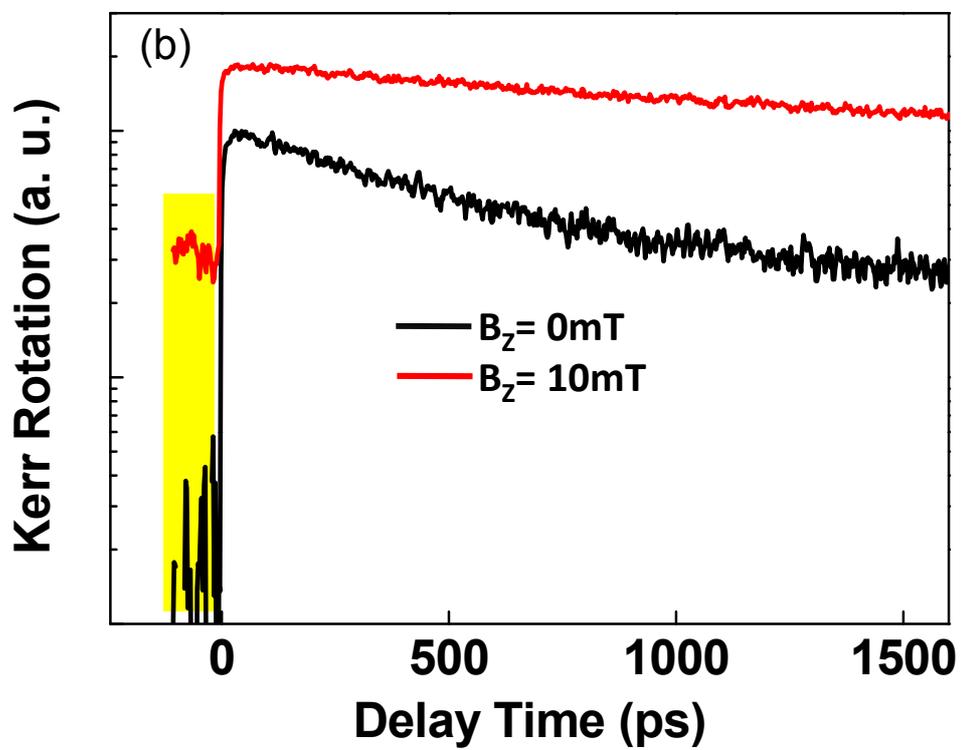

Figure 2

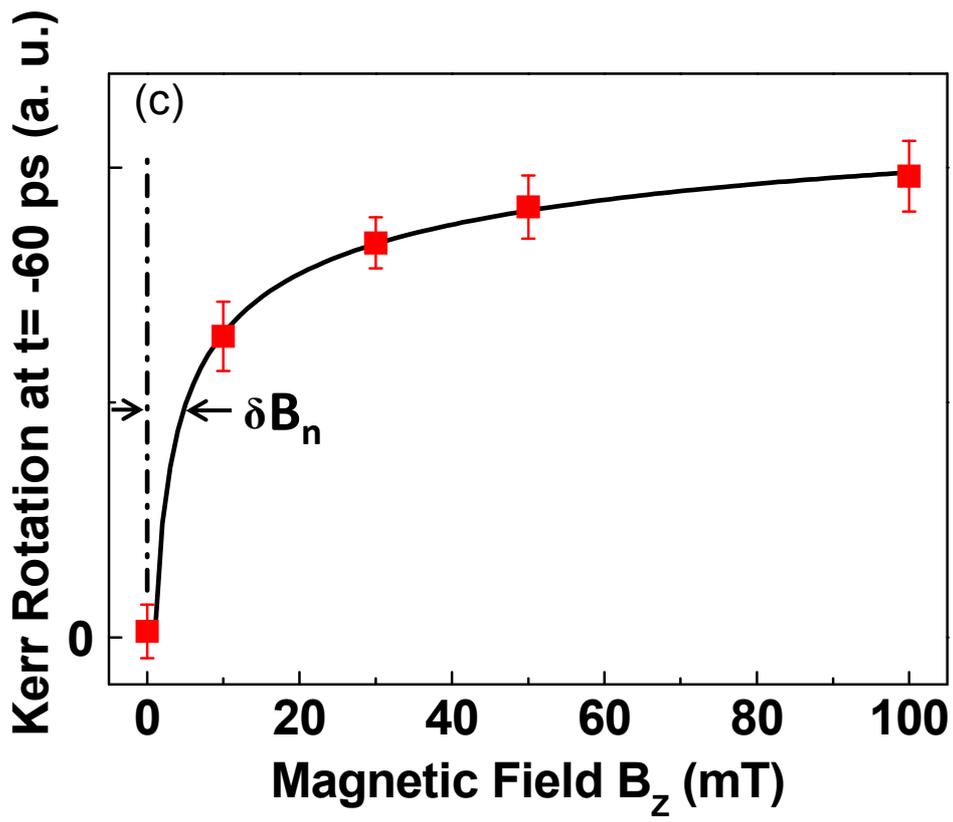

**Figure 2**